\documentclass[cits]{PoS}
\usepackage[authoryear,square]{natbib}
\bibpunct{(}{)}{;}{a}{}{,}

\title{Pulsar Science with the SKA}

\ShortTitle{Pulsars and the SKA}

\author{
\speaker{Michael Kramer}$^{1,2}$, 
Ben Stappers$^2$
\\ 
$^1$MPI f\"ur Radioastronomie, Auf dem H\"ugel 69, 53121 Bonn, Germany\\
$^2$Jodrell Bank Centre for Astrophysics, University of Manchester, M13
9PL, UK
\\
E-mail: \email{mkramer at mpifr.de}
}

\abstract{ The SKA will be transformational for many areas of science,
  but in particular for the study of neutron stars and their usage as
  tools for fundamental physics in the form of radio pulsars. Since
  the last science case for the SKA, numerous and unexpected advances
  have been made broadening the science goals even further. With the
  design of SKA Phase 1 being finalised, it is time to confront the
  new knowledge in this field, with the prospects promised by this
  exciting new telescope. While technically challenging, we can build
  our expectations on recent discoveries and technical developments
  that have reinforced our previous science goals.}

\FullConference{
Advancing Astrophysics with the Square Kilometre Array\\
June 8-13, 2014\\
Giardini Naxos, Sicily, Italy}

\newcommand{\skipthis}[1]{}

\begin{document}

\section{Introduction}

Pulsars are a physicist's dream come true. Their study combines a wide
range of physics and astrophysics, from the fundamental laws of nature to
the structure of the Milky Way. Pulsars, as compact objects of the most
extreme matter, store a huge amount of rotational energy, making them
massive flywheels and very stable rotators. With a radio beam fixed
relative to the surface of this rotating object, which has a strange
superfluid superconducting material inside a solid iron crust, it acts like
a cosmic lighthouse. On long time-scales, the frequency stability of the
lighthouse's beacon rivals atomic clocks on Earth and allows us to test the
terrestrial time standards. If the pulsar has a companion, they fall
together in the gravitational potential of the Galaxy. By studying the way
in which they fall, we can test general relativity and alternative theories
of gravity under strong-field conditions in a unique and elegant, simple
and clean way by measuring the times-of-arrival (ToA) of the pulses on
Earth. These ToAs and the pulse properties are modified by the interstellar
medium, in turn allowing us to probe its properties. As we are always
looking for even better laboratories, we stare at the sky with our radio
telescopes to find new pulsars with ever better sensitivity and
instrumentation. Once in a while, we discover something truly unexpected,
such as the Fast Radio Bursts (FRB) from cosmological distances. With the
Square Kilometre Array (SKA) and its sensitivity, we will not only be able
to continue this research but we will take it to a new level. This overview
chapter summarises the specialised chapters in the remainder of the book
and puts them into perspective of what the community proposed in 2004 in
the chapters by Kramer et al. (2004) and Cordes et al. (2004) in the SKA
Science Book edited by Carilli \& Rawlings (2004).

\subsection{The discovery potential of the SKA}

\begin{figure}[h]
\vspace{-20pt}
\begin{minipage}{8cm}
  % lbrt -> brtl with the +90 rotation
  \includegraphics[scale=0.38,angle=90,trim=50mm 20mm 20mm 50mm, clip]{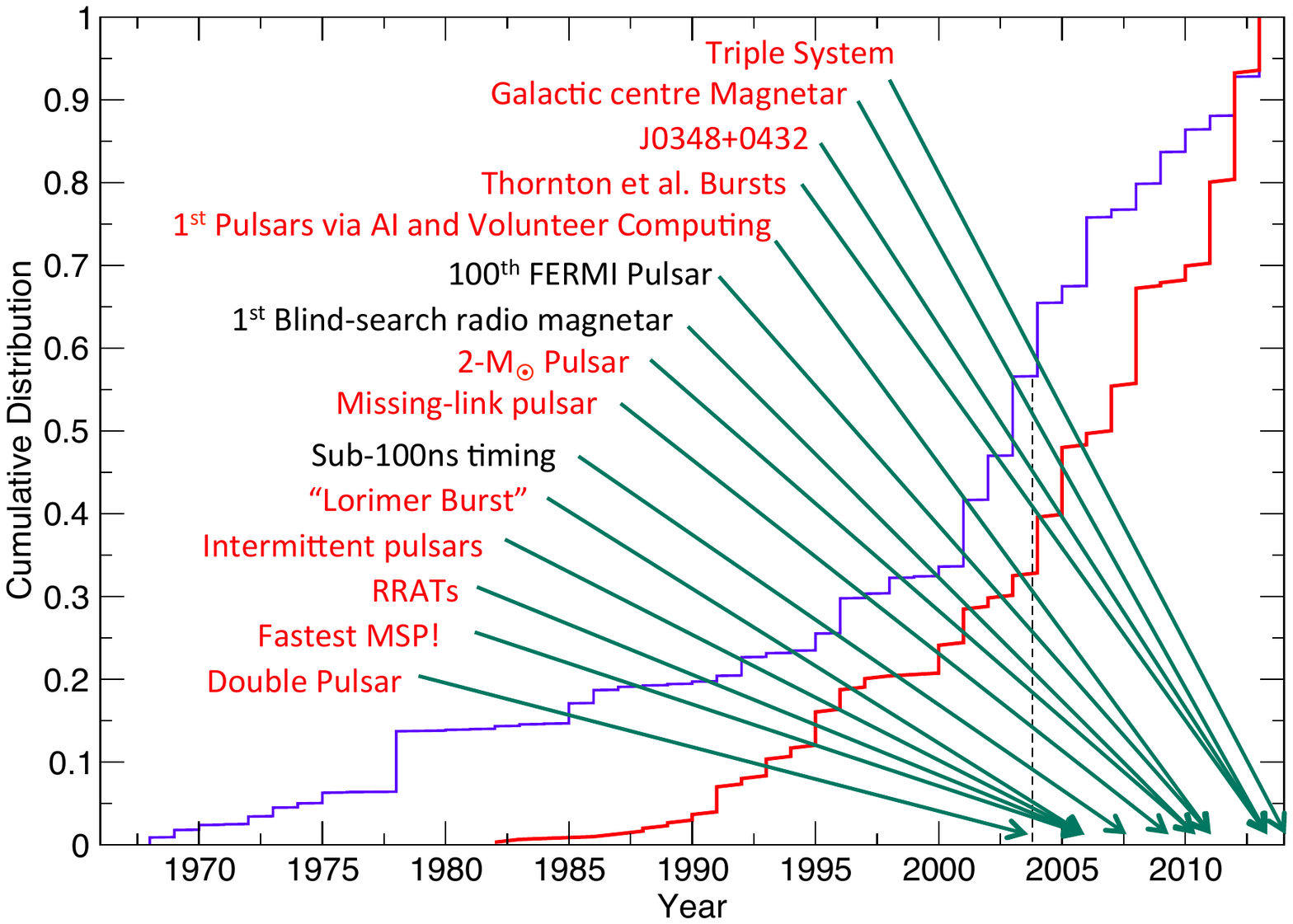}
\end{minipage}
\begin{minipage}{8cm}
%\vspace{-0.4cm}
  % lbrt -> tlbr with the -90 rotation
  \includegraphics[scale=0.38,angle=-90,trim=0mm 45mm 30mm 0mm, clip]{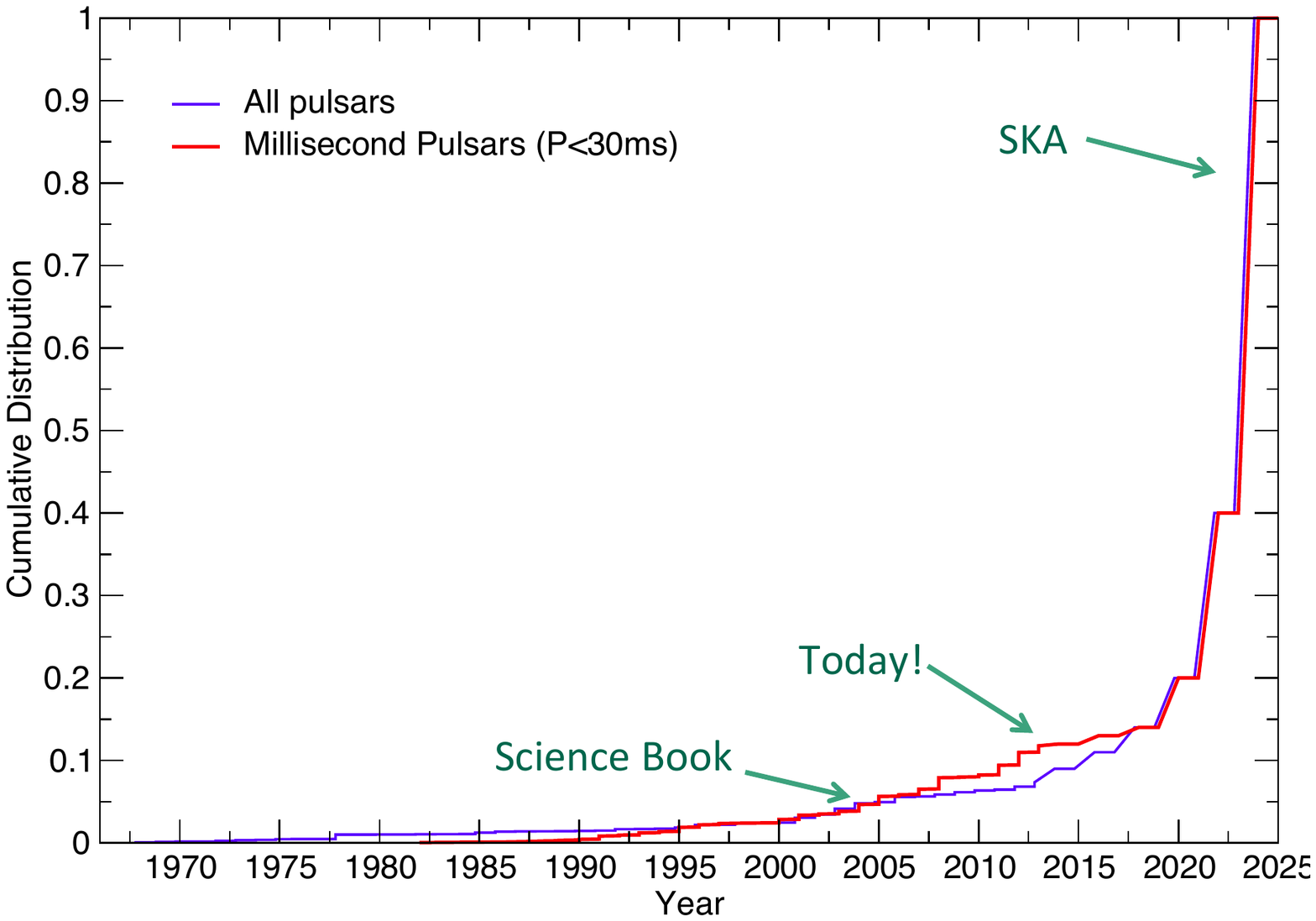}
\end{minipage}
  \caption{Pulsar-related discoveries as a function of time. The time
    of the first SKA Science Book is marked and some important
    (selected) discoveries since are marked. The right panel puts the
    current numbers into perspective with those expected for the SKA.
  }\label{fig:summary}
\end{figure}

Almost all the science presented in the ``pulsar chapters'' in this book
and summarised below was already proposed for the SKA in 2004. However that
definitely does not mean that, in the intervening 10 years, this research
field has stood still. The very opposite is true. If anything, the rate of
discoveries has even increased, largely due to the employment of new
instrumentation, providing unprecedented sensitivity and time resolution --
all possible due to the recent ability to digitise and process large
bandwidths with commodity computing equipment.  Far exceeding all of this,
the SKA will provide yet another gigantic leap in various areas of
instrumentation, giving orders of magnitude improvement in sensitivity,
FoV, survey speed and many more. In order to demonstrate this, we summarise
the relevant discoveries in Figure~\ref{fig:summary}.  The left panel shows
the cumulative increase in pulsar discoveries as a function of time, with a
number of important discoveries in the last decade. In total, we know of
more than 2300 normal pulsars and about 250 millisecond (recycled) pulsars,
of which about 80\% are in binaries. With the full SKA we can expect more
than a 10-fold increase in each of these numbers, a large fraction of which
can be already found with Phase I, as described in the census chapter by
Keane et al. (2015). These numbers are even somewhat larger than predicted
by Kramer et al. (2004) and Cordes et al. (2004) as our knowledge of the
populations has increased dramatically. The right panel of
Figure~\ref{fig:summary}  puts the expected numbers in relation to the
current ones, demonstrating that these new discoveries will lead clearly to
excellent science, resulting from the subsequent (needed) follow-up and
timing observations, as described in the accompanying chapters.

\section{Science enabled by the discovery \& study of pulsars and radio emitting neutron stars with the SKA}

%few sentence summary, perhaps with some highlight plots for selected
%chapters, not sure about the order yet - SHall we put in context with
%other efforts with other instruments? May be the most useful bit
%rather than repeating everything

The pulsar key science described in the first SKA Science Book had a number
of related components, which were summarised under the theme of ``Testing
Gravity''. With pulsars being strongly self-gravitating bodies and
precision clocks at the same time, timing observations of binary and
isolated millisecond pulsars allow unprecedented strong-field experiments.
These include testing general relativity and alternative theories of
gravity using binary pulsars and (the yet to be discovered) pulsar-black
hole systems as well as the direct detection of gravitational waves using a
``Pulsar Timing Array'' (PTA) experiment. Given the advances in recent
years, prospects are now described in two separate chapters by Shao et al.
(2015) and Janssen et al. (2015), respectively.  In addition to those, we
provide here a summary of the rich and varied science goals for the SKA
described in the appropriate chapters:

%\begin{description}
\textbf{Chapter 37 --- Gravitational wave astronomy with the SKA ---
  Janssen et al. (2015)} A Pulsar Timing Array (PTA) is used as a cosmic
gravitational wave (GW) detector. As described in the chapter by
Janssen et al. (2015), Phase I essentially guarantees the direct detection of
a GW signal. This may appear as a stochastic background from binary
super-massive black holes in the process of early galaxy evolution, or
it may be bright individual source(s) of this kind. Exotic phenomena
like cosmic strings may also be expected to produce measurable GW
signals, should they exist. The last ten years have seen a much better
understanding of the source population, the detection procedures and
the use of a PTA for fundamental physics (such as graviton properties,
e.g.~\citealt{ljp+10}) or single source localisation capabilities
(e.g.~\citealt{lwk+11}), all of which is described in the
corresponding chapter.

\textbf{Chapter 38 --- Understanding pulsar magnetospheres with the
SKA --- Karastergiou et al. (2015)} Considerable progress has been made
with our understanding of the pulsar emission mechanism in the last
decade. However, the wide bandwidth and exceptional sensitivity of the
SKA will revolutionise our understanding of radio emission from all
types of radio emitting neutron stars.  Combined with the excellent
high frequency data sets being accumulated by instruments like XMM,
Chandra, FERMI, Magic, Veritas and HESS and new data from
contemporaneous instruments like CTA, the data from SKA in Phase~1
will allow us to map the magnetosphere with unprecedented detail. With
the SKA we will have an even greater range of radio frequencies
available and sensitivity to study with more detail a greater number
of sources. The high cadence of observing that should be afforded by
the wide fields of view, sub-arraying capabilities, multi-beaming and
commensal observing modes will revolutionise our understanding of the
crucial relationship between the emission and spin properties of large
samples of pulsars.  The raw sensitivity of both MID and LOW will not
only improve our understanding of the magnetospheric phenomena such as
nulling and moding in a significantly increased sample of pulsars, it
will also reveal in more detail how these manifest in millisecond
pulsars. Both of these elements will be crucial in understanding the
pulsar emission physics, but also for improving the use of pulsars as
precision timing instruments. As we go from Phase~1 to SKA there will
be significant improvements in the number of sources that can be
studied, due to the large number of new sources and increased
sensitivity, and even wider range of frequencies while the improved
computing will enable even higher cadence.

\textbf{Chapter 39 - Understanding the Neutron Star Population with
  the SKA - Tauris et al. (2015)} The census will give a complete overview of
the Galactic neutron star population. Not only has the number of
neutron stars increased significantly in the last 10 years, but also
new types of neutron stars have been discovered, such as Intermittent
Pulsars \citep{klo+06} or RRATs \citep{mll+06}, and unexpected types
of binary systems. Also, some magnetars are now detected at radio
frequencies, and it is therefore essential that we revisit the birth
channels and properties of neutron stars and their evolution. The
ingredients needed to understand the complex ``zoo'' of neutron stars
and their inter-relationships is summarised in the chapter by Tauris
et al. (2015).

\textbf{Chapter 40 - A Cosmic Census of Radio Pulsars with the SKA -
  Keane et al. (2015)} To transform and facilitate all of the science
discussed here and in this book, it is essential to discover the
largest possible fraction of the entire Galactic radio-emitting
neutron star population. This is possible with the SKA. Moreover, this
population, in itself, will provide an unparalleled opportunity to
study the population as a whole and all that can tell us about the
star formation and dynamical history of the Milky Way.  This chapter
presents a comprehensive analysis of the survey approaches and
expected yields. We refer here only to selected highlights, noting
that in the sample of radio emitting neutron stars we include RRATs
and magnetars. In Phase~1 of the SKA, our simulations show that an
optimised combination of pulsar search capabilities using both LOW and
Bands 1 \& 2 of MID can survey the entire sky visible from the SKA
sites in a reasonable amount of observing time.  Such a survey, with
the rebaselined configurations of MID and LOW, would result in a yield
of at least 7500 normal pulsars and 1200 millisecond pulsars. This
assumes similar integration times on MID and LOW, if there were more
time available on LOW a doubling of the integration time could be used
to increase this yield by as much as 30\%. Full SKA simulations have
shown that a combination of a LOW, MID and an Aperture Array
instrument working at below a GHz, could increase the total number of
radio emitting neutron stars known in the Galaxy to more than 45,000
including more than 5,000 millisecond pulsars. In the majority of the
regions of the Galaxy this would constitute the entire radio emitting
neutron star population.  In the Galactic centre a deep dedicated
survey with Band 5 of the first phase of SKA will be sensitive to
normal and millisecond pulsars, with the high frequency mitigating the
deleterious effects of the interstellar medium (ISM) and the
collecting area compensating for the steep pulsar spectra.  With more
than 1,000 pulsars predicted to be located with in the central parsec,
in Phase 1 we can expect to detect dozens of pulsars in this region of
the sky, and the later addition to the SKA of bands 3 \& 4 combined
with overall improved collecting area, will discover the vast majority
of this predicted population.

\textbf{Chapter 41 - Three-dimensional Tomography of the Galactic and
  Extragalactic Magnetoionic Medium with the SKA - Han et al. (2015)}
Another spin-off from the pulsar surveys will be the significant
increase in the number of lines-of-sight through the Galaxy where we
will be able to measure both the magnetic field properties and the
electron column density, see this chapter for a full discussion. In
Phase~1 there will already be an order of magnitude increase in the
number of sources and the combination of LOW and MID will allow both
near and distant sources, including some deep in the centre of the
Galaxy, to be studied. The wide bandwidths, high frequency resolution
and large number of sources will provide the accurate rotation
measures and dispersions measure to generate a 3D tomographic view of
the magneto-ionic properties of the Galaxy. With the full SKA
revealing almost the entire radio pulsar population in the Galaxy it
will allow one to probe the far side and also study the finer
detail. Already in Phase~1 we will be able to find pulsars outside of
our Galaxy and so study the fields inside those galaxies and also in
the intervening medium. However the real renaissance in this area will
come with the exceptional sensitivity of the SKA.

\textbf{Chapter 42 - Testing Gravity with Pulsars in the SKA Era -
  Shao et al. (2015)} New discoveries also lead to new exotic laboratories
for fundamental physics, in particular for the study of
gravity. Discoveries in recent years have continued to confirm this
impression. The unique Double Pulsar system was found unexpectedly
\citep{bdp+03,lbk+04}, providing unique and, so far, the best tests of
general relativity for strongly-self gravitating bodies
\citep{ksm+06}; the most massive neutron star in a relativistic binary
was found and probes a previously untested regime for alternative
theories of gravity \citep{afw+13}; and the SKA observations of a
unique discovered triple system \citep{rsa+14} promises to enable
tests of the Strong Equivalence Principle that will surpass solar
system limits.

We expect that among the discoveries are also pulsar-black hole
systems. Using pulsars as test masses, we can probe the properties of
black holes and measure the mass and the spin, $\chi$. This will lead
to precision tests of the ``Cosmic Censorship Conjecture'', which
states that every astrophysical black hole should have an event
horizon (cf.~\citealt{lewk14}). In order to test also the ``no-hair''
theorem, stating that all black hole properties are uniquely described
by the mass and spin (and charge) of the black hole, we would need to
also measure the quadrupole moment, $Q$, of the black hole, which
should be identical to the value predicted by the measured combination
of mass and spin.

\textbf{Chapter 43 - Probing the neutron star interior and the
  Equation of State of cold dense matter with the SKA - Watts et al. (2015)}
Finding a large population of neutron stars will also yield the
extreme objects. The SKA surveys will yield the fastest spinning
pulsars, perhaps even sub-ms pulsars should they exist, the most
massive pulsars and also of high interest, the lightest neutron
stars. All this information will be used to study the
equation-of-state (EOS) of super-dense matter. With neutron stars
consisting of the densest matter in the observable Universe, we cannot
create the corresponding conditions in terrestrial laboratories, so
that pulsar observations, including those of glitches,
moment-of-inertia measurements, provide a unique insight into the
EOS(s?), as discussed in detailed in the chapter by Watts et
al. Combining this insight with X-ray observations will also be
extremely valuable as discussed further below.
%\item Chapter 44 - ???? - Does not exist!

\textbf{Chapter 45 - Observing Radio Pulsars in the Galactic Centre
  with the Square Kilometre Array - Eatough et al. (2015)} Measuring the
quadrupole moment $Q$, and hence testing the ``no-hair'' theorem, for
stellar mass black holes is difficult, even with the SKA. It will be
much easier using a pulsar in orbit about the super-massive black hole
in the centre of our Milky Way. The recipe for how to extract this
science using observations of relativistic effects such as
frame-dragging was first presented by \citet{wk99}. This was further
studied \citep{kbc+04,lwk+12} and is discussed in much detail in the
corresponding chapter by Eatough et al. (2015). The SKA prospects of measuring
the mass of SGR A* to a precision of $1M_\odot$ (!), its spin $\chi$
to 0.1\% or better and the quadrupole moment $Q$ to about 1\%, even
with a normal pulsar orbiting SGR A* is exciting and unprecedented.

\textbf{Chapter 46 - Pulsar Wind Nebulae in the SKA era - Gelfand et
  al.} As well as making a great step forward in studying pulsars and
their environments using their properties in the time domain, the SKA
can also make enormous strides by studying the environments of pulsars
through imaging. This chapter discusses how the bulk of the rotational
energy of pulsars is radiated away in the pulsar wind and the only way
to study it is through its interaction with the surrounding
interstellar medium or with a binary companion and/or its wind. The
SKA, using both the MID and LOW elements, in Phase 1 will already
unveil dozens of new pulsar wind nebulae providing vital new insight
in to the composition of the wind and the acceleration of leptons up
to PeV energies. These interactions will also reveal important details
of the ISM, the companion star composition and their winds. When
combined with the existing and upcoming data from instruments such as
HESS, Veritas, Magic and Fermi and that from instruments such as CTA,
this will revolutionise our understanding of pulsar winds and their
environments.

\textbf{Chapter 47 - Pulsars in Globular Clusters with the SKA -
  Hessels et al. (2015)} The sheer number of pulsars and radio emitting
neutron stars in globular clusters make them ideal targets for the
SKA, especially as the vast majority are best viewed from the Southern
Hemisphere. As discussed in this chapter, the excellent sensitivity of
both LOW and MID in Phase~1 will already more than triple the number
of pulsars known in globular clusters, while the SKA will discover a
sample of millisecond pulsars to rival the rest of the Galactic
population. Not only does this provide the opportunity to find rare
systems for some of the applications discussed above, it provides
unique tools to probe the magneto-ionic properties of the clusters,
the star formation and dynamical history of these important Galactic
building blocks. In the vein of targeted searches both LOW and MID
will be excellent instruments for searching for pulsars in sources
that are likely to contain neutron stars, either through their high
frequency emission, in particular Fermi unidentified sources, or their
steep spectrum radio emission. The ability to frequently search these
targets will also reveal more of the evolutionarily important
transition sources, that mark the birth of radio millisecond
pulsars. Already in Phase 1 we will be to find dozens more pulsars in
the Magellanic clouds enabling us to sample the pulsar population in
completely different environments to our own Galaxy and we will sample
the top end of the luminosity function of any pulsars located in other
nearby dwarf galaxies. By looking for giant pulse like emission, as
exhibited by the Crab pulsar, the reach will extend well out into the
local group galaxies in Phase~1. With the SKA we will not only sample
a very large fraction of the pulsars in the Magellanic clouds but will
be able to detect significant numbers of pulsars in many of the
galaxies in the local group and giant pulse emitters out to the Virgo
Cluster.

\textbf{Chapter 157 - SKA and the next-generation multi-wavelength
  observatories - Antoniadis et al. (2015)} We have already alluded to a few
of the synergies between pulsar work with the SKA and other
instruments. However there are many more, and these are detailed in
this chapter on synergies. These include the synergy with new and next
generation gravitational wave detectors where the discovery of an
increased number of double-neutron star binaries will enhance our
understanding of the binary merger rate, the discovery of thousands of
more radio pulsars may unveil some that are sufficiently distorted to
emit gravitational waves, provide timing ephemerides for all of these
pulsars to search for gravitational waves. There is also the
possibility of the PTAs uncovering burst sources of GWs that can be
followed up with instruments like LSST, IXO/Athena, Astro-H and of
course proposed binary supermassive black holes can be searched for in
the PTA data.  Combining the astrometric information on binary systems
with optical studies of their companions with instruments like LSST,
E-ELT and GAIA can provide new insight into the stellar evolution,
Galactic dynamics and cooling properties and ages of the
stars. Combined X-ray and radio searches can reveal new transition
systems which can reveal more about the formation process of
millisecond pulsars and studies at both wavelengths can provide
important input into our understanding of the neutron star equation of
state.  In particular, we are looking forward to synergies with
Advanced LIGO and other ground (or even space-based) detectors. One
one hand, they complement the SKA tests in the strong-gravity regime
in a nearly perfect way, while they also allow us to probe similar
sources in different evolutionary stages and across the mass scale.
%\end{description}

\begin{figure}[t]
%\centerline{
%\psfig{file=plot4.ps,width=13cm} 
%}
  \centering
  \includegraphics[width=13cm]{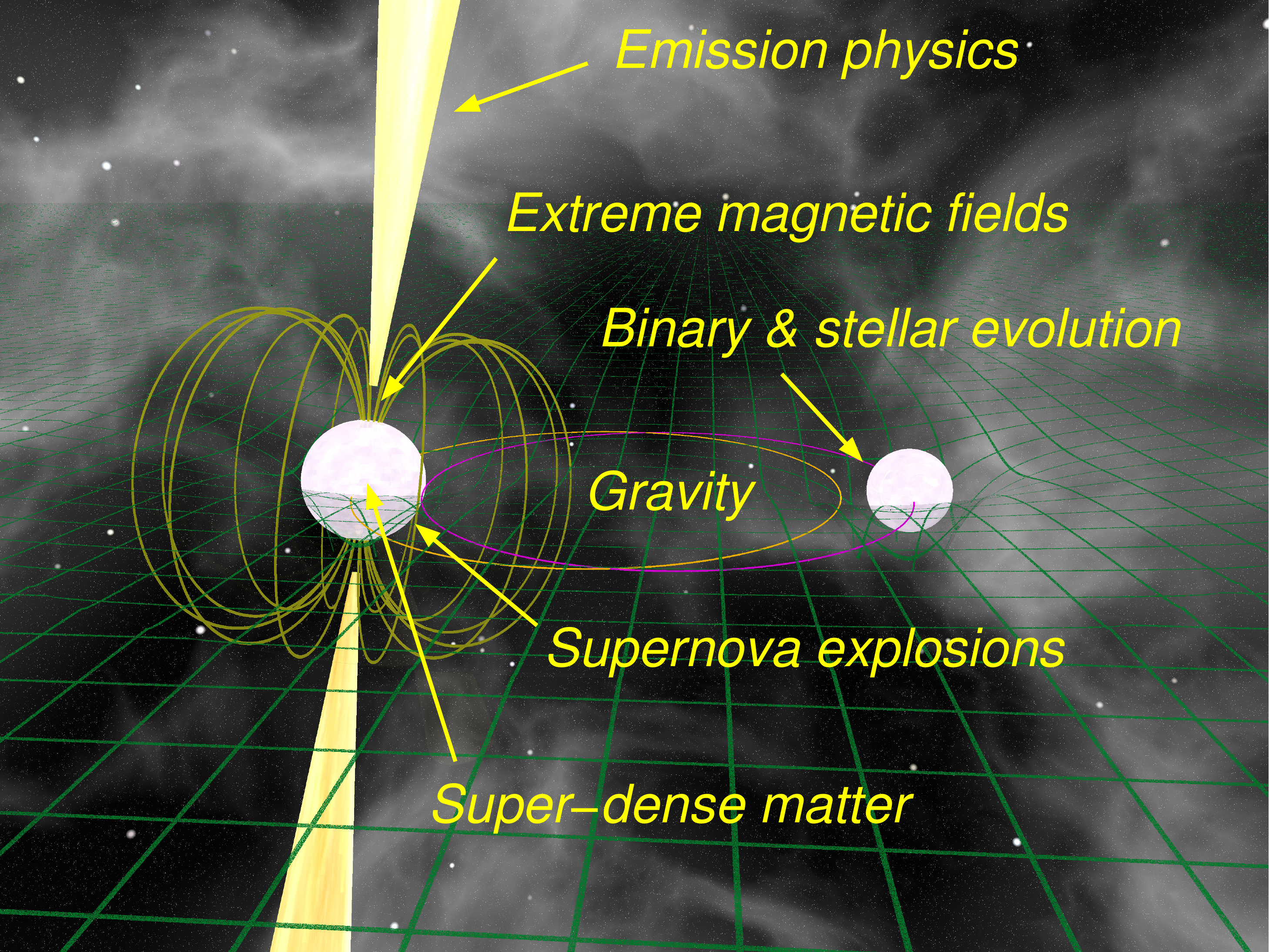}
  \caption{Artistic impression demonstrating the wide range of physics
    and astrophysics that finds its application when studying pulsars.}
  \label{fig:synergies}
\end{figure}

\section{Where does the SKA fit?}

\subsection{The synergy between SKA LOW and MID}

It is important to recognise that the combined value of SKA LOW and
MID in both Phase~1 and Phase~2 is more than the sum of its
parts. This is true for all of the pulsar science discussed in this
book, except only for the study of the innermost regions of the
Galaxy.  The significant increase in computing power to allow the
removal of dispersion in the ISM combined with the ultra-wide
bandwidths of instruments like LOFAR, MWA and the LWA have highlighted
the value of returning to the lowest few octaves of the radio spectrum
visible from the Earth for both pulsar searching and detailed pulsar
studies. The excellent sensitivity of LOW already in Phase~1 will make
it the ideal instrument for performing pulsar searches off the plane,
where it is likely to find at least half of the known pulsars beamed
in our direction. On the other hand the higher frequencies of MID will
allow us to reach deep into the Galactic plane and see through the ISM
to find thousands of pulsars there. It is essential though, that for
both instruments there are sufficient numbers of tied-array beams
available to ensure a sufficiently rapid survey speed with
sufficiently long integration times. This will be possible
with a minimum of 500 tied-array beams available for LOW and about
1500 beams for MID in Phase~1.

The ability of the SKA to have almost continuous frequency coverage
from 50\,MHz up to about 15\,GHz with continuous up to 1.8\,GHz in
Phase~1, is of real value for studies of the pulsar emission
mechanism and magnetospheres, providing the ability to sufficiently
sample the variations in emission properties as a function of
frequency.  The ability to correct for the influence of the ISM on
high precision timing will be essential for all of the gravity studies
and this is made possible through measuring the ISM weather through
the continuous frequency bands. This is even more important if the
highest frequency for pulsar timing in Phase 1 is around
1.8\,GHz. Studies of the ISM and the Galactic distribution of
electrons and the structure of the magnetic fields also value the
combination of frequencies allowing measurements to be made nearby
where the effects are best seen at LOW frequencies, and further away
using the higher frequency bands. In the case of the pulsar wind
nebulae the two instruments provide an ability to trace the energy
deposition history and the physics of the pulsar wind.

\subsection{Synergies with other SKA science goals}

The pulsar surveys and associated observing and processing
infrastructure naturally lead to a strong overlap with the fast
transients goals (see Chapter 51). It is highly desirable therefore
that the infrastructure, including the beamformer, be made available
for searching for short duration radio bursts, be they from radio
emitting neutron stars or from other radio transients, operate
commensally, ideally with all other observations with both SKA MID and
LOW, but certainly as frequently as possible. This will increase the
yield for both the pulsar and radio emitting neutron star searches but
also for the fast transients. As discussed in Chapter 41 and others,
pulsars are very useful for the study of the ISM, providing full
knowledge about the Milky Way, its constituents and its magnetic field
structure. Synergies will also be present by comparing experimental
results in gravity (via pulsars and cosmology) and the detailed
characterisation of individual sources. In general, the aim should be
to share discoveries with these communities as soon as possible to
maximise the science return as rapidly as possible.

\subsection{Impact of the SKA}

The impact of the pulsar science obtained with the SKA promises to be
very significant. Pulsars are among the very few objects that are
visible across the whole electromagnetic spectrum and through new,
unexplored windows like gravitational wave astronomy. At the same
time, the understanding requires the application of theories ranging
from quantum mechanics and solid states physics to gravitation and
extreme plasma physics. Consequently, the variety of expected results
is guaranteed to be relevant for a global community of physicists and
astronomers alike. Literally, we can expect that we can mark our
understanding of neutron stars, gravity and the Milky Way itself in a
time before and after the SKA came online.

Besides this massive scientific impact, the pulsar science with the
SKA also has the potential to draw great attention and interest from
the general public. Neutron stars and in particular the black holes
studied with the pulsars always fascinate the public. The fact that we
can visualize pusar data in a unique dynamic way --- and make them
even audible --- will make it easy to advocate the SKA and its science
in general.

\subsection{Risk and challenges}

The technological components of ``non-imaging processing'' are
challenging. This involves the task of beamforming and, due to the
high data rates, the required on-line search processing. Data can not
be stored but have to be analysed in real time. Imperfect algorithms
would require a repeat of the experiment or parts of it. This may be
both expensive or even impossible. On the other hand, the discovery 
rate could be high, consequently. 

There are also scientific risks that are more difficult to quantify,
as we are conducting discovery science. The biggest risk here is
probably the uncertainty in the number of suitable pulsar - black hole
systems or of pulsars about Sgr A* to conduct the proposed key
science. Nevertheless, we know that extreme binary systems can be
found in globular cluster (see Chapter~47) or in the regions of high
stellar density in the central Galaxy. Here, the discovery of a rare
radio-loud magnetar \cite{efk+13} however gives confidence that the
expections are warranted (see Chapter~45).

\section{Technical requirements of non-imaging processing with the SKA}

The key technical requirements of the SKA are beyond the scope of this
chapter and so we refer the interested reader to the papers by Smits
et al. (2009, 2011) \nocite{slk+09,stw+11} as well as to the detailed
technical design from both the Non-Imaging work in the Central Signal
Processing and Science Data Processor
consortia\footnote{www.skatelescope.org} and also in the associated
Chapters in this book. However, we highlight here some of the key
requirements on both SKA MID and LOW to achieve the pulsar science
goals.

\begin{itemize}
\item It is essential that in Phase 1 of SKA MID and LOW that there be a
sufficiently large number of beams (e.g. 1,500 and 500 respectively) to be
able to survey the sky sufficiently rapidly and with sufficient sensitivity. 
\item When searching for the extreme binary systems that are key to the gravity studies
it is not possible to trade integration time for sensitivity as the required processing capacity scales
as the third power of the integration time. So maximum sensitivity possible in the core
regions needs to be maintained. 
\item Sub-arraying and a significant number of beams capable of undertaking precision timing are required
to be able to enable characterisation of the large numbers of pulsars that will be discovered, but also to
enable the high cadence observing required to reveal the physics of neutron star interiors and magnetospheres. 
\item  Commensal observing will allow for transformation changes to  the observing cadence achievable on the
largest possible sample of objects. 
\item It should be possible to combine as many of the elements as possible, when forming the smaller number of 
beams to be used for the timing experiments to maximise sensitivity. 
\item Having as complete as possible frequency coverage across the
  entire LOW range and the ability to subarray so that that different
  frequencies can be observed simultaneously will be highly valuable
  for the physics of the magnetosphere.
\item A combination of pulsar timing on both the LOW and MID instruments is required in order to be able to fully
characterise and correct for the contribution of the ISM in the high precision pulsar timing experiments. 
\item It will be necessary to be able to calibrate the polarisation purity of the array to at least -40 dB for high precision timing experiments. 
\end{itemize}

%The SKA is an interferometer with sparsely distributed elements, as is 
%required for obtaining images of sufficient spatial resolution for many of
%the science goals of the instrument. On the timescale of the SKA it will
%not be possible to generate and process the typical interferometer output, 
%that is visibilities, at sufficient high time resolution for pulsar searching. 
%It is therefore necessary to find a compromise whereby a core area of 
%the SKA is defined within which there is a higher spatial density of receiving 
%elements which allows for significant sensitivity and lower spatial resolution. Instead
%of forming the correlated products from these elements they are usually combined
%to form beams, which typically sample a region of sky smaller than the primary
%beam described by the individual elements, simply because it isn't possible 
%to process all of these beams. For a more detailed discussion of these issues 
%see Smits et al ????

\section{Summary}

The science summarized in this overview can only give a
glimpse of the science promises given by the SKA. With Phase 1 of
the telescope being defined, we can safely predict what will be
possible. However, the most exciting results will be, of course, obtained by
unpredicted, unexpected discoveries and results. We may not be able to
imagine what awaits us, but even the ``expected'' results are
strong motivation to make the SKA a reality.

%\bibliographystyle{mn}   % if natbib is available 
%\bibliography{journals,psrrefs,psrrefs_add.bib,modrefs,crossrefs}

\begin{thebibliography}{90}
\expandafter\ifx\csname natexlab\endcsname\relax\def\natexlab#1{#1}\fi

\bibitem[{Antoniadis {et~al.}(2013)}]{afw+13} Antoniadis K. {et~al.},
  2013, Science, 340, 448

\bibitem[{Antoniadis {et~al.}(2015) Antoniadis, J., Guillemot, L.,
    Possenti, A. et. al.}]{ant14} Antoniadis, J., Guillemot, L.,
  Possenti, A. et al., 2015, ``Multi-wavelength, Multi-Messenger
  Pulsar Science in the SKA Era'', in proc. {\em Advancing
    Astrophysics with the Square Kilometre Array}, PoS(AASKA14)157

\bibitem[{Burgay {et~al.}(2003)}]{bdp+03} {Burgay}~M. {et~al.},
  2003, Nature, 426, 531

\bibitem[{CarilliRawlings (2004)}]{cr04}
Carilli, C., \& Rawlings, S., 2004, eds. ``Science with the Square
Kilometre Array", NewAR, Vol. 48

\bibitem[{Cordes {et~al.}(2004)}]{ckl+03}
Cordes, J. M., Kramer, M., Lazio, T. J. W., Stappers, B. W., Backer, D. C.,
Johnston, S., 2004, NewAR, 48, 1413

\bibitem[{Eatough {et~al.}(2013)}]{efk+13} {Eatough}~R.~P. {\rm
  et~al.}, 2013, Nature, 501, 391

\bibitem[{Eatough {et~al.}(2015) Eatough, R. P., Lazio, J. T. W.,
    Casanellas, J. et. al.}]{eat14} Eatough, R. P., Lazio, J. T. W.,
  Casanellas, J. et al., 2015, ``Observing Radio Pulsars in the
  Galactic Centre with the Square Kilometre Array'', in proc. {\em
    Advancing Astrophysics with the Square Kilometre Array},
  PoS(AASKA14)045

\bibitem[{Fender {et~al.}(2015) Fender, R., Stewart, A., Macquart,
    J-P. et. al.}]{fen15} Fender, R., Stewart, A., Macquart, J-P. et
  al., 2015, ``Transients with the SKA: the scientific potential, and
  how to optimise the telescope'', in proc. {\em Advancing
    Astrophysics with the Square Kilometre Array}, PoS(AASKA14)051

\bibitem[{Gelfand {et~al.}(2015) Gelfand, J. D., Breton, R. P., Ng,
    C.-Y. et al.}]{gel15}Gelfand, J. D., Breton, R. P., Ng, C.-Y. et
  al., 2015, ``Pulsar Wind Nebulae in the SKA era'', in proc. {\em
    Advancing Astrophysics with the Square Kilometre Array},
  PoS(AASKA14)046

\bibitem[{Han {et~al.}(2015) Han J. L., van Straten, W., Lazio,
    T. J. W. et. al.}]{han14} Han J. L., van Straten, W., Lazio,
  T. J. W. et. al., 2015, ``Three-dimensional Tomography of the
  Galactic and Extragalactic Magnetoionic Medium with the SKA'', in
  proc. {\em Advancing Astrophysics with the Square Kilometre Array},
  PoS(AASKA14)041

\bibitem[{Hessels {et~al.}(2015) Hessels, J. W. T., Possenti, A.,
    Bailes, M. et. al.}]{hes14} Hessels, J. W. T., Possenti, A.,
  Bailes, M. et al., 2015, ``Pulsars in Globular Clusters with the
  SKA'', in proc. {\em Advancing Astrophysics with the Square
    Kilometre Array}, PoS(AASKA14)047

\bibitem[{Janssen {et~al.}(2015) Janssen, G. H., Hobbs, G.,
    McLaughlin, M. A. et. al.}]{gemma14} Janssen, G. H., Hobbs, G.,
  McLaughlin, M. A. et al., 2015, ``Gravitational Wave Astronomy with
  the SKA'', in proc. {\em Advancing Astrophysics with the Square
    Kilometre Array}, PoS(AASKA14)037

\bibitem[{Karastergiou {et~al.}(2015) Karastergiou, A., Johnston, S.,
    Andersson, N., {et al.}}]{ch38} Karastergiou, A., Johnston, S.,
  Andersson, N., {et al.}, 2015, ``Understanding pulsar magnetospheres
  with the SKA'', in proc. {\em Advancing Astrophysics with the
    Square Kilometre Array}, PoS(AASKA14)038

\bibitem[{{Keane} {et~al.}(2015){Keane}, {Bhattacharyya}, {Kramer}, \&
    {et~al.}}]{kbk+14} {Keane}, E.~F., {Bhattacharyya}, B., {Kramer},
  M., \& {et~al.} 2015, ``A Cosmic\ Census of Radio Pulsars with the
  SKA'', in proc. {\em Advancing Astrophysics with the Square
    Kilometre Array}, PoS(AASKA14)040

\bibitem[{Kramer {et~al.}(2004)}]{kbc+04} Kramer~M., Backer~D.~C.,
  Cordes~J.~M., Lazio~T. J.~W., Stappers~B.~W., Johnston~S., 2004,
  NewAR, 48, 993

\bibitem[{Kramer {et~al.}(2006)}]{klo+06} {Kramer}~M., {Lyne}~A.~G.,
  {O'Brien}~J.~T., {Jordan}~C.~A., {Lorimer}~D.~R., 2006, Science,
  312, 549

\bibitem[{Kramer {et~al.}(2006)}]{ksm+06} {Kramer}~M. {\rm et~al.},
  2006, Science, 314, 97

\bibitem[{Lee {et~al.}(2010)}]{ljp+10} {Lee}~K., {Jenet}~F.~A.,
  {Price}~R.~H., {Wex}~N., {Kramer}~M., 2010, ApJ, 722, 1589

\bibitem[{Lee {et~al.}(2011)}]{lwk+11} {Lee}~K.~J., {Wex}~N.,
  {Kramer}~M., {Stappers}~B.~W., {Bassa}~C.~G., {Janssen}~G.~H.,
  {Karuppusamy}~R., {Smits}~R., 2011, MNRAS, 414, 3251

\bibitem[{Liu {et~al.}(2012)}]{lwk+12} {Liu}~K., {Wex}~N.,
  {Kramer}~M., {Cordes}~J.~M., {Lazio}~T.~J.~W., 2012, ApJ, 747, 1

\bibitem[{Liu {et~al.}(2014)}]{lewk14} {Liu}~K., {Eatough}~R.~P.,
  {Wex}~N., {Kramer}~M., 2014, MNRAS, 445, 3115

\bibitem[{Lyne {et~al.}(2004)}]{lbk+04} Lyne~A.~G. {\rm et~al.}, 2004,
  Science, 303, 1153

\bibitem[{McLaughlin {et~al.}(2006)}]{mll+06} {McLaughlin}~M.~A. {\rm
  et~al.}, 2006, Nature, 439, 817

\bibitem[{Ransom {et~al.}(2014)}]{rsa+14} {Ransom}~S.~M. {\rm et~al.},
  2014, Nature, 505, 520

\bibitem[{{Shao} {et~al.}(2015){Shao}, {Stairs}, {Antoniadis}, \&
{et~al.}}]{ssa+14} 
{Shao}, L., {Stairs}, I.~H., {Antoniadis}, J., \&
{et~al.}, 2015, ``Testing Gravity with Pulsars in the SKA Era'', in proc.
{\em Advancing Astrophysics with the Square Kilometre
Array}, PoS(AASKA14)042

\bibitem[{Smits {et~al.}(2009)}]{slk+09} {Smits}~R., {Lorimer}~D.~R.,
  {Kramer}~M., {Manchester}~R., {Stappers}~B., {Jin}~C.~J.,
  {Nan}~R.~D., {Li}~D., 2009, A\&A, 505, 919

\bibitem[{Smits {et~al.}(2011)}]{stw+11} {Smits}~R., {Tingay}~T.,
  {Wex}~N., {Kramer}~M., {Stappers}~B., 2011, A\&A, 528, 108

\bibitem[{Tauris {et~al.}(2015) Tauris, T.~M., Kaspi, V.~M., Breton,
    R.~P. et al.}]{tkb+14} Tauris, T.~M., Kaspi, V.~M. Breton,
  R.~P. et~al., 2015, ``Understanding the Neutron Star Population with
  the SKA'', in proc. {\em Advancing Astrophysics with the Square
    Kilometre Array}, PoS(AASKA14)039

\bibitem[{Watts {et~al.}(2015) Watts, A. L, Xu, R., Espinoza, C. M. et
    al.}]{anna14} Watts, A. L, Xu, R., Espinoza, C. M. et al. 2015,
  ``Probing the neutron star interior and the Equation of State of              
  cold dense matter with the SKA'', in proc. {\em Advancing                     
    Astrophysics with the Square Kilometre Array}, PoS(AASKA14)043

\bibitem[{Wex {et~al.}(1999)}]{wk99} Wex~N., Kopeikin~S., 1999, ApJ,
  513, 388

\end{thebibliography}

\setlength{\bibsep}{0pt}

%\begin{thebibliography}{99}
%
%\bibitem[1]{}
%Carilli, C.~L., \& Rawlings, S.\ 2004, \nar, 48, 979 
%
%\bibitem[2]{} Ekers, R.\ 2012, arXiv:1212.3497
%
%\bibitem[3]{} Ekers, R.\ 2012, \apj
%
%\end{thebibliography}

\end{document}